\documentclass[12pt]{iopart}
\usepackage{graphicx}

\begin{document}

\title{Fast rotation of nuclei with extreme isospin in the
vicinity of neutron and proton drip lines.}

\author{A. V. Afanasjev$^1$, S. Teeti$^1$ and A. Taninah$^2$}

\address{$^1$Department of Physics and Astronomy, Mississippi State University,
Starkville, MS 39762, USA}

\address{$^2$Department of Chemistry and Physics, Louisiana State University, 
Shreveport, Louisiana 71115, USA}

\ead{aa242@msstate.edu}
\vspace{10pt}
%\begin{indented}
%\item[]August 2017
%\end{indented}

\begin{abstract}
The analysis of the present understanding of collective rotation in very neutron-rich
nuclei is presented. It is shown that collective rotation can lead to the increase of
stability of rotational states with increasing spin. The detailed investigation of 
rotational excitations in very proton-rich nuclei confirms this conclusion and indicate
that experimental studies of such features are more feasible in the nuclei near
proton drip line. They also show that rotational bands which are proton quasi-bound
at  zero or low spins can be transformed into proton bound ones at high spin by collective 
rotation of nuclear  systems.  This is due to strong Coriolis interaction which acts on 
high-$j$ or strongly mixed M orbitals and drives the highest in energy occupied single-particle 
states into negative energy domain. These physical mechanisms lead to a substantial 
extension of the nuclear landscape beyond the spin zero proton drip line.  In  addition, 
a new phenomenon of the formation of giant proton halos in rotating nuclei emerges:  
it is triggered by the occupation of strongly mixed M intruder  orbitals. 
\end{abstract}

\maketitle

%\maketitle

%%%%%%%%%%%%%%%%%%%
\section{Introduction}
%%%%%%%%%%%%%%%%%%%

   The rotation is one of the most important collective excitations in atomic nuclei. 
It leads  to  a lot of interesting physical phenomena such as multiply rotational bands 
in a specific nucleus based on different particle-hole excitations \cite{Szy-book,VDS.83,NilRag-book}, 
superdeformation \cite{Dy152-first-SD.86,BHN.95,VALR.05}, gradual evolution of collectivity in 
rotational  bands up  to their final termination in a single-particle terminating states \cite{PhysRep-SBT}, 
magnetic and anti-magnetic rotational bands \cite{F-rev.01,Meng2013Front.Phys.55},
clusterization \cite{IMIO.11,ZIM.15,AA.18} etc. However, with a few exceptions experimental 
and theoretical  studies of rotational excitations have been carried out for nuclei located not far 
away from the $\beta$-stability valley and more frequently on its proton-rich side in the direction of 
proton-drip line. On the other hand, the introduction of new experimental facilities such as FRIB will 
potentially extend the studies of rotational structures to significantly more proton- and neutron-rich 
nuclei as compared with those investigated till now. 

    In the nuclei experimentally investigated so far the rotational excitations reveal themselves
via the rotational bands the rotational states of which are connected by $\gamma$-transitions
of different multipolarity i.e. by the E2 transitions in the rotational bands which have an electric
character and by the M1 transitions in magnetic bands 
\cite{NilRag-book,PhysRep-SBT,F-rev.01,Meng2013Front.Phys.55}.  The 
rotational states are discrete because all occupied single-particle states  in the nucleonic
configurations of the bands are particle bound: typical $\gamma$-decay time $\tau_{\gamma}$
is of the order of $10^{-9}$ s. 

 The situation is expected to be different in the nuclei located near the drip lines since the last occupied 
singe-particle state(s) of nucleonic configuration can be located at positive energies. This will 
lead to an opening of other channels of the decay such as proton and neutron emission competing 
with the $\gamma$-decay from rotational states. However, only few papers have been dedicated
to theoretical studies of rotational motion in the nuclei near neutron drip line \cite{NDMMS.01,FNJMP.16,FRMLN.16,AIR.19} and no 
experimental studies for such systems exist. The situation is opposite for the nuclei near the 
proton drip line: the rotational bands based on ground state proton-emitting configurations 
have been experimentally observed in the $^{59}$Cu \cite{59Cu-pm.02} and  $^{141}$Ho 
\cite{141Ho.01} nuclei but theoretical studies of rotational motion in the nuclei located near proton
drip line have not been carried out.  Thus,  the goals of the present paper are the following:
(i) to review existing theoretical studies and to extract the general features of rotational
motion in the nuclei located near the drip lines and (2) to carry out a detailed investigation of rotational
structures in the nuclei located near the proton drip line.

  The paper is organized as follows. An overview of theoretical studies of rotational
structures in the nuclei near neutron drip line is presented in Sec.\ \ref{n-drip-status}.
Open questions related to rotational motion in the nuclei located near the 
drip lines and their resolution are discussed in Sec.\ \ref{open-q}. The general features of the rotation
in proton-rich nuclei and the differences with the situation near  the neutron-drip line are
briefly overviewed in Sec.\ \ref{rot-prot}. The details of theoretical calculations are presented
in Sec.\ \ref{theory}.  The birth of proton-bound  rotational bands is illustrated in Sec.\ \ref{Birth-sec}
on the example of the $^{14}$Ne and $^{67}$Kr nuclei. The emergence  of giant proton halos 
at very high spin is discussed in Sec.\ \ref{Halo-sect}. Rotating induced extension of the
nuclear landscape is considered in Sec.\ \ref{Extension-sect}. Finally, Sec.\ \ref{concl} summarizes 
the results of our  paper.

%%%%%%%%%%%%%%%%%%%%%%%%%%%%%%%%%%%%%%%%%
\section{Rotation in neutron-rich nuclei near neutron-drip line: the current status}
\label{n-drip-status}
%%%%%%%%%%%%%%%%%%%%%%%%%%%%%%%%%%%%%%%%%

   At present, there are no experimental studies of rotational bands in the neutron-rich nuclei
located at or near neutron drip line. However, several theoretical studies reveal interesting 
features of rotational excitations in such nuclei which are distinctly different from those seen in 
known nuclei. They emerge from the possibility of neutron emission. Note that nucleonic configuration 
is a system of nucleons distributed over specific set of single-particle states  which is bound 
together for a time $\tau_{conf}$ which is substantially larger than characteristic nuclear timescale  
of  $10^{-22}$s (see Ref.\ \cite{Thoen-review.04}). Note that setting up collective rotation 
also requires time which is substantially larger than above mentioned timescale. Thus, the 
rotational  states cannot be built on the occupation of the single-particle states which belong to 
the  scattering continuum (i.e. from the states with orbital momentum $l\neq 0$ located at the 
energies larger than $E_B$  (see Fig.\ \ref{scheme}(a) and Ref.\  \cite{Thoen-review.04})
or from positive energy single-particle states with $l=0$ (see Refs.\ \cite{FNJMP.16,FRMLN.16}). 

%%%%%%%%%%%%%%%%%%%%%%%%%%%%%%%%%%%%%%%%
\begin{figure*}[ht]
\centering
\includegraphics*[width=11.5cm]{fig-1-rev.eps} 
\caption{A schematic representation of nuclear bound, quasi-bound (resonant) 
and scattering states. Here, $V_{eff}(r)=V_{nucl} + V_{cfg} + V_{Cou}$ represents effective 
potential formed by nucleonic (central) $V_{nucl}$, centrifugal $V_{cfg}$ and Coulomb 
$V_{Cou}$ potentials. Note that the Coulomb potential is absent in the neutron subsystem. 
$E_B$ is the height of the barrier of the effective potential the peak of which is located in 
near-surface region of the nucleus. Decaying quasi-bound states are located at the energies 
$e_i$ in the range of $0 < e_i <E_B$. See text for further details.
\label{scheme}
}
\end{figure*}
%%%%%%%%%%%%%%%%%%%%%%%%%%%%%%%%%%%%%%%%

  Nucleonic configurations in very neutron-rich nuclei which lead to rotational states are 
built from the combination of bound $|B>$ and quasi-bound $|QB>$ states (see Fig.\ 
\ref{scheme}(a)) and the properties of occupied $|QB>$ state(s) define the stability of configuration
with respect of neutron emission. Note that the lifetimes for neutron emission from the ground 
and low-spin states in the $A \leq 28$ nuclei located near neutron-drip line, for which experimental 
data exist, are very short: due to low value of centrifugal barrier  the largest half life is only by two 
orders of magnitude larger than $10^{-22}$ s \cite{DG.23}.  Even in this kind of situation, it is 
possible in very neutron-rich $^{11}$Be and $^{39}$Mg nuclei to build 
rotational bands formed  on the basis of the nucleonic configurations in which the occupied 
$|QB>$ state has orbital momentum $l\neq 0$ \cite{FNJMP.16,FRMLN.16}. The rotational
states of such bands have a width because of coupling with continuum \cite{FNJMP.16,FRMLN.16}
and their observation is possible only via neutron emission which proceeds substantially
faster than the typical $\gamma$-decay. 

   The rotational bands calculated in these two very neutron-rich nuclei have two interesting 
features.  First, the states of favored in energy signature branch of the ground state $1/2^+$ 
rotational band in $^{11}$Be are predicted to have small neutron decay width of around 200 
keV, while these widths are substantially larger (around 0.7 MeV) for the states in unfavored 
in energy signature branch \cite{FNJMP.16}. This means that unfavored signature states are 
more unbound (i.e. located at higher energy and thus facing lower effective barrier with 
smaller width) than favored ones. The same features exist in our cranked relativistic mean
field (CRMF) calculations: the ground state rotational band in $^{11}$Be is based on the 
deformed 1/2[220] Nilsson state emerging from the $d_{5/2}$ spherical neutron subshell and 
there is sufficient signature splitting between favored and unfavored signature branches to 
account for the difference in decay width. Second, the calculations of Ref.\ \cite{FRMLN.16} 
for rotational  band in $^{39}$Mg based on deformed $1/2[321]$ Nilsson state show drastic 
decrease of the decay width with increasing angular momentum in the band from very broad 
to very narrow resonances. This is due to the fact that the structure of odd neutron in low 
and middle spin $1/2^-$, $3/2^-$, $5/2^-$, $7/2^-$ rotational states contains substantial 
contribution from the $l=1$ $p_{3/2}$ spherical subshell while the one for high-spin $9/2^-$ 
and $11/2^-$ states is dominated by the $l=3$ $f_{7/2}$ states.  As a consequence, the 
drastic reduction of the decay width at high spin is due to the substantial increase of 
centrifugal barrier.

  Note that the calculations of rotational structures in Refs.\ \cite{FNJMP.16,FRMLN.16} have 
been carried within  the particle-plus-core model based on a nonadiabatic coupled-channel 
formalism and the Berggren single-particle ensemble. The advantage of this model is that it 
takes explicitly into account bound states, narrow resonances, and the scattering continuum.
However, it uses the core fitted to existing low spin data in neighboring even-even nucleus 
and  assumes  fixed deformation for the single-particle Woods-Saxon potential which does 
not depend on angular  momentum of rotational states. However, the deformation of rotational 
states (especially in light nuclei) is expected to change (drastically) with increasing angular 
momentum (see, for example, Refs.\ \cite{NilRag-book,PhysRep-SBT,CEMPRRZ.95,RA.16} 
and references  quoted therein).

   This deficiency of the particle-plus-core model can be taken care in the cranking model
which defines the deformation of a rotational state of interest in a fully self-consistent way.
However, before our investigations there was only one exploratory study of weakly-bound 
neutron-rich even-even $^{30-38}$Ne and $^{32-40}$Mg nuclei within  the cranked 
Skyrme-Hartree-Fock (CSHF) approach without pairing in Ref.\ \cite{NDMMS.01}. The deformed 
configurations considered in this paper have the valence neutrons in weakly-bound 
$f_{7/2}$  high-$j$ intruder states, located at small negative energies. They have large 
centrifugal barrier due to large orbital angular momentum $l=3$ which leads to a good 
localization of their wavefunctions (see Ref.\ \cite{NDMMS.01}).

%%%%%%%%%%%%%%%%%%%%%%%%
\section{Open questions and their resolution}
\label{open-q}
%%%%%%%%%%%%%%%%%%%%%%%%

   However, there are additional factors at play not considered in these studies.
For example, they examine only low spin yrast or near yrast configurations 
with relatively modest deformation. However, the angular momentum which can be built 
within such rotational bands is quite limited. Let us take as an example the $^{38}$Ne nucleus. 
The lowest configuration in this nucleus has the 
$\pi (d_{5/2})^2_4 \otimes \nu (core)_0$ structure with respect of the $(Z=8, N=28)$ core
with maximum angular 
momentum $I_{max}=4^+$  and the 
deformed configuration built by two particle - two hole excitation across
the $N=28$ spherical shell gap has the structure 
$\pi(d_{5/2})^2_4 \otimes \nu (f_{7/2})^{-2}_6 (f_{5/2})^2_4$
with maximum angular momentum $I_{max}=14^+$. Note that the configurations
are labelled here in the language of Ref.\ \cite{PhysRep-SBT}. Other examples of limited angular 
momentum in the rotational structures built on the ground or excited configurations of light nuclei
of interest are discussed in Sec. 12.4 of Ref.\ \cite{NilRag-book} (for example, ground state rotational band 
of $^{20}$Ne terminates at  $I_{max}=8^+$), Sec. 7 of Ref.\ \cite{PhysRep-SBT} and Sec. III of Ref.\ 
\cite{RA.16}.  Thus, this analysis leads to {\it a first question}, namely, {\it "What is the mechanism 
of the creation of higher  angular momentum in the nuclei of interest and whether such 
mechanism leads to a destruction or stabilization of rotational motion in a given nucleonic
configuration?"}
  
    Absolute majority of rotational bands studied so far both in experiment and in theory are 
built on nucleonic configurations which are particle bound, i.e. all occupied single-particle 
states of these configurations are located at negative energies.  Thus, appropriate name
for this class of rotational bands is {\it particle-bound rotational bands.} The second class
of rotational bands, in which at least one occupied single-particle state of respective 
nucleonic configuration is located at positive energy below $E_B$ (i.e. the state is 
quasi-bound \cite{Sakurai-QM,Chong-QM-III-book,SSN.97}\footnote{Quasi-bound 
states are denoted also as quasistationary or resonance states in the literature (see, for 
example, Ref.\ \cite{SSN.97}).}), has been suggested in Refs.\ \cite{FNJMP.16,FRMLN.16}. There is no well 
established name for this class of bands: they were called as rotational bands embedded 
in the continuum in Ref.\ \cite{FRMLN.16} and as resonance bands in Ref.\ \cite{AIR.19}.
In the present paper we call them as {\it quasi-bound rotational bands} to stress that
their states are bound for sufficiently long time to form a collective structure and that 
they can decay by particle emission.  This discussion raises the {\it second 
and third questions}, namely, {\it "Can these two classes of rotational bands coexist 
in the same nucleus?"} and {\it "Whether the transition from one class of collective rotation 
to another is possible within a rotational band built on a fixed nucleonic configuration?"}.

 %%%%%%%%%%%%%%%%%%%%%%%%%%%%%%%%%%%%%%%%%%%
\begin{figure}[ht]
\centering
\includegraphics*[width=9.5cm]{fig-2-a-rev.eps}
\includegraphics*[width=9.5cm]{fig-2-b-rev.eps}
\caption{Proton single-particle energies (routhians) in the self-consistent rotating 
potential of $^{14}$Ne as a function of rotational frequency $\Omega$. They are 
given along the deformation path of the $[1,1,1,1] \otimes [2,3,3,2]$ [panel (a)]
and  $[1,1,1,1] \otimes [1,3,4,2]$ [panel (b)] occupation  blocks (see Sec.\ \ref{conf-label}
for more details). Long-dashed  red, 
solid black, dot-dashed green, and dotted blue lines 
indicate $(\pi = +, r = +i)$, $(\pi = +, r = -i)$,
$(\pi = -, r = +i)$, and $(\pi = -, r = -i)$ orbitals, respectively. At $\Omega = 0.0$ MeV,
the single-particle orbitals are labeled by the asymptotic quantum numbers $K[Nn_z\Lambda]$ 
(Nilsson quantum numbers) of the dominant component of the wavefunction. Solid circles indicate 
occupied orbitals in proton unbound and proton bound parts of  the configurations.  The vertical 
orange dashed line indicates the frequency (corresponding to spin $I_{trans}$) at which the 
configuration becomes proton bound. Intruder orbitals of the M-type before and after band 
crossing are labelled by I1 and I2 in panel (a). See text for further details.
\label{routh}
}
\end{figure}
%%%%%%%%%%%%%%%%%%%%%%%%%%%%%%%%%%%%%%%%%%%%%%
  
   To build higher angular momentum than the one in the ground state rotational 
band one should excite the particles from the orbitals with low orbital angular 
momentum $l$ into the ones with high $l$ values (see Refs.\ 
\cite{NilRag-book,PhysRep-SBT}). Thus, the high spin configurations which are 
either yrast or located close to the yrast line are expected to be characterized by 
the occupation of the single-particle orbitals with high  values of orbital momentum $l$ 
(see Ref.\ \cite{AIR.19}). This has interesting consequences for the nuclei 
near drip lines since the $l=0$ and $l=1$ orbitals govern halo properties and particle 
decay. The centrifugal barrier (see Fig.\ \ref{scheme}) increases substantially with the 
increase of $l$ and this leads both to a suppression of halos and particle decays 
and to a better localization of the wave function of the $l>1$ orbitals 
within the nuclear volume \cite{NDMMS.01}.

   This situation can be illustrated by the analysis of the single-particle routhian diagram 
shown in   Fig.\ \ref{routh}  which is obtained in the CRMF calculations. Note that the
slope $\partial e_i^{\Omega}/\partial \Omega$ of the single-particle routhian 
$e_i^{\Omega}$ shown in this figure indicates its orbital angular momentum $l$.  
This is because the value of the single-particle alignment 
$\left< j_x \right> = - \partial e_i^{\Omega}/\partial \Omega$ of the orbital
is defined  by its orbital angular momentum (see Fig. 1 in Ref.\ \cite{Pd101}). 
Note also that high-$l$ orbitals become almost fully aligned with the axis of
rotation at relatively low frequencies.  By considering the $0-4.0$ MeV energy range
of Fig.\ \ref{routh} one can conclude that  on average with increasing rotational 
frequency the slopes of the single-particle routhians (and, as a result, their angular 
momentum $l$)  become larger and larger.  With increasing rotational frequency the 
routhians located in the vicinity of the Fermi level  dive faster under the effective 
barrier of Fig.\ \ref{scheme} leading to the situation in which particle emission 
proceeds at the single-particle energies $e_i^{\Omega}$ characterized by a higher 
value of $E_B-e_i^{\Omega}$ and the width of the barrier at $e_i^{\Omega}$. 

This analysis suggests that even if the configurations remain quasi-bound
the combination of these two factors can lead to an increase of the stability
of rotational states with respect of particle emission i.e. to a reduction of decay
width with increasing spin. This result is in line with those for rotational 
$1/2[321]$ band in $^{39}$Mg obtained within particle-plus-core model in 
Ref.\ \cite{FRMLN.16} which show drastic decrease of the decay width
(from very broad to very narrow resonances) with increasing angular momentum 
in the band.

 To address the second question, let us assume that the routhian diagram presented
in Fig.\ \ref{routh}(b) is also valid for deformed configurations in the $^{12}$O nucleus 
which has two protons less than $^{14}$Ne. Under this assumption, the highest occupied 
proton single-particle orbitals in $^{12}$O are 1/2[101]: the occupation of the
single-particle orbitals in this configuration is show by open squares in Fig.\ \ref{routh}(b).
The rotational band built on this configuration will be weakly bound up to rotational frequency 
$\Omega \approx 2.0$ MeV and its description in the CRMF framework is the same as for 
weakly bound configurations in neutron-rich nuclei near the neutron drip line discussed in 
the CSHF approach in Ref.\ \cite{NDMMS.01}.  However, the excitations [shown by green 
dashed arrows in Fig.\ \ref{routh}(b)] of proton from the $1/2[101](r=+i)$ orbital into either 
the $1/2[220](r=-i)$ or $3/2[211](r=-i)$ ones lead to the excited configurations in which the 
proton in the highest occupied orbital is quasi-bound.  The rotational bands built on these
two configurations are quasi-bound from $\Omega=0.0$ MeV up to $\Omega \approx 3.0$ MeV.
Although this is schematic illustration, fully self-consistent CRMF  calculations based
on systematic survey of configurations of interest in the nuclei near proton and neutron drip lines
confirm this picture and indicate that particle-bound and quasi-bound rotational bands can 
coexist in the same nucleus. This typically takes place when the ground  state rotational 
band is weakly bound in the nucleus located near drip line.

  To address the third question let us consider the evolution of the single-particle routhians 
in the $[1,1,1,1] \otimes [2,3,3,2]$ occupation block with increasing rotational frequency 
shown in Fig.\ \ref{routh}(a). The last occupied $3/2[211](r=-i)$ orbital in this configuration is 
quasi-bound and is characterized by a substantial slope which indicates moderate value of 
orbital angular momentum $l$.  An unpaired band crossing between this orbital and the
I2 one takes place at rotational frequency $\Omega \approx 2.6$ MeV.  The I2 orbital is
significantly more downsloping with increasing rotational frequency and thus it has  
substantially larger orbital angular momentum
than the  $3/2[211](r=-i)$ one.  As a consequence, with increasing rotational frequency
it becomes bound at $\Omega \approx 3.2$ MeV. Thus, the rotational sequence built
on such configuration would exhibit the properties of quasi-bound rotational band
at $\Omega < 3.2$ MeV and particle-bound rotational band at higher rotational frequency. 
Such kind of transition (denoted as a birth of particle-bound rotational  bands in Ref.\ 
\cite{AIR.19})  has been investigated for the first time in neutron-rich nuclei near the neutron 
drip line in Ref.\ \cite{AIR.19} but as illustrated in the present paper it is active also in 
very proton-rich nuclei at and beyond spin-zero proton-drip line. Alternative possibility of the transition 
from particle-bound to quasi-bound rotational band (the death of particle-bound rotational bands 
in the language of Ref.\ \cite{AIR.19})  with increasing spin also exists but it is less frequent in 
the calculations.  Thus, the present analysis clearly indicates that particle-bound and quasi-bound
rotational  bands do not exist as strictly separate classes of  the bands: the transition between 
them is possible even within fixed nucleonic configuration.

Note that as collective coordinates the rotational frequency $\Omega$ in rotating nuclei 
and deformation $\beta_i$ in non-rotating ones can have similar effect on binding of 
the nucleons. The former case is discussed above. The transformation of the character of
single-particle orbitals from unbound/quasi-bound(resonant) to bound and vise versa 
under the change of deformation are clearly visible in the Nilsson diagrams presented
in Refs.\ \cite{Ham.07,Ham.19}. 
  
   The present analysis clearly illustrates that the building of the angular 
momentum higher than the one which can be achieved in the ground state 
rotational band requires particle-hole excitations from the states with low orbital 
momentum $l$ to the ones with high $l$ values.  In the nuclei near drip lines,
the latter are typically located at positive energies. Thus, these excitations lead 
to an increase of centrifugal barrier and, as a consequence, to a significant 
reduction of the decay width of the rotational states of quasi-bound rotational  
bands.  Moreover, with increasing rotational frequency Coriolis interaction acting 
on occupied high-$j$ (high-$l$) orbitals, which are quasi-bound at low frequencies,
drives them below zero energy threshold  and makes them bound (see also 
Ref.\ \cite{AIR.19}). Thus, one can conclude that through these two mechanisms 
the collective rotation can lead to an increase of the stability of nuclear systems. 
Note that the rotational structures at the spin larger than that generated
in the ground state rotational bands of the very neutron-rich nuclei have been 
systematically studied only in Ref.\ \cite{AIR.19} in the CRMF framework.  In 
particular,  it was shown that the latter mechanism can lead to an extension
of nuclear landscape to higher neutron numbers as compared with that obtained
at low spin.

%%%%%%%%%%%%%%%%%%%%%%%%%%%%%%%%%%
\section{Fast rotation in proton-rich nuclei: general features}
\label{rot-prot}
%%%%%%%%%%%%%%%%%%%%%%%%%%%%%%%%%%
   
   The main focus of the present study is fast rotation in very proton rich 
nuclei. In such nuclei the situation is drastically different as compared 
with very neutron rich ones because of the presence of Coulomb 
barrier [see Fig.\ \ref{scheme}(b)] which has a substantial value. For example, 
the Coulomb barrier (at spherical shape) almost linearly increases from 2 to 6 MeV
in the $N=Z$ nuclei when $Z$ changes from 10 to 36. In combination with 
centrifugal barrier [see Fig.\ \ref{scheme}(b)]  it leads to a drastic increase (by many orders of magnitude) 
of proton emission half-lives as compared with characteristic nuclear timescale 
of $10^{-22}$s \cite{SSN.97,WD-rev.97,DLW.06-PRL}. For  instance, the present
window  for the experimental observation of proton  emission half-lives in medium 
mass nuclei ranges from $4.5 \times 10^{-7}$ s to a few seconds which 
corresponds to extremely narrow width  $\Gamma \approx 10^{-22} - 10^{-15}$ MeV of 
respective proton resonances \cite{SSN.97}.  In that respect these quasi-bound proton 
states are  very similar to the bound ones. With decreasing proton number and 
lowering the Coulomb barrier the proton resonances are expected 
to become broader but for typical situations of interest they are
anticipated to be substantially narrower than the neutron ones discussed
earlier. Note that the main focus  of this paper is on proton bound parts of rotational 
bands in which the highest in energy occupied single-particle states are located at 
negative energies.

%%%%%%%%%%%%%%%%%%%%%%%%%%%
\section{The details of theoretical calculations}
\label{theory}
%%%%%%%%%%%%%%%%%%%%%%%%%%%

%%%%%%%%%%%%%%%%%%%%%%%%
\subsection{Theoretical framework}
\label{theory}
%%%%%%%%%%%%%%%%%%%%%%%%

   The calculations are performed in the framework of cranked relativistic mean field 
(CRMF) theory \cite{VALR.05,KR.89}.  It represents the  realization of  covariant 
density functional theory for rotating nuclei with no pairing correlations in one-dimensional 
cranking approximation.  It has been successfully tested in a systematic way on the 
properties of different types of rotational bands in the regime of weak pairing such as 
normal-deformed, superdeformed and smooth terminating bands as well as the bands 
at the extremes of angular momentum (see Refs.\ \cite{VALR.05,AF.05} and references 
therein). We restrict ourselves to reflection symmetric shapes which are dominant deformed 
shapes  in the nuclear chart.

  The cranking model is semiclassical in nature (see Sec. 3.4 of Ref.\ \cite{RS.80}) 
the microscopic derivation of which is provided in projected approaches (see Sec. 11 
in Ref.\ \cite{RS.80}): the cranking model appear as a limit of these approaches at 
large deformation. It has a proven track of successful description of rotational structures 
of different types such as normal- and superdeformed rotational bands, terminating bands, 
magnetic and antimagnetic bands across the nuclear chart (see Refs.\ 
\cite{Szy-book,VDS.83,NilRag-book,VALR.05,PhysRep-SBT,F-rev.01,Meng2013Front.Phys.55,BF.79}
and references quoted therein). In reality,  the absolute majority (more than 90\% of the cases) 
of  experimental rotational bands, for which theoretical descriptions  exist, have been successfully 
analyzed in the framework of the cranking model. Note that the rotational structures discussed 
below have a deformation  which is large enough  for a cranking model to be valid.  Moreover  
the equivalence of the description of rotating nuclei in the intrinsic (rotating) and laboratory frames 
was confirmed with good accuracy in Ref.\ \cite{CEMPRRZ.95} on the example of $^{48}$Cr. 
In addition,  earlier investigations of the nuclei in the regions of the nuclear chart located near
those studied in the present paper show that cranking model successfully describes experimental 
data on rotational bands (see Refs.\ \cite{CEMPRRZ.95,RA.16,AF.05,A60}) and provides
interesting predictions for the structure of rotational bands in light nuclei (see Refs.\ 
\cite{IMIO.11,ZIM.15,RA.16}).  Note that similar to cranking model for rotating nuclei, 
the transformation to the intrinsic rotating frame is also the most dominant approach in the study 
of rotating Bose-Einstein condensates (see Ref.\ \cite{Fetter.09}) and rotating quark-gluon 
plasma  (see Refs.\ \cite{JLL.16,CG.17,EFM.17}).

   Available studies of high spin properties of rotating $N\approx Z$ nuclei 
located in the vicinity of the part of nuclear chart under study clearly indicate that static 
pairing becomes negligible at high spin  \cite{RA.16,AF.05}.   This allows safely 
neglect the pairing correlations in the calculations since the focus of the present study 
is on the parts of rotational bands which become proton bound at high spin.
The spin at which the calculations without pairing become a good approximation 
to those with pairing depends on the mass region: it is around $I\approx 15\hbar$  in the Kr 
($Z\approx 36$)  region \cite{AF.05} but  decreases with lowering the proton number 
$Z$ \cite{RA.16} down to approximately $5\hbar$ in the nuclei with $Z\leq 10$. In addition, 
the pairing is substantially reduced due to blocking in the nucleonic configurations in which 
the opposite signature orbitals are not pairwise occupied  \cite{VALR.05,RS.80}: for such 
configurations the calculations without pairing reproduce  experimental data even at lower 
spins \cite{AF.05}.

  The CRMF calculations are carried out with the NL3* covariant energy density  
functional (CEDF)  \cite{NL3*} which is the state-of-the-art functional for nonlinear 
meson-nucleon coupling model \cite{AARR.14}. It is globally tested for ground state 
observables in even-even nuclei \cite{AARR.14}.   The CRMF and cranked relativistic 
Hartree-Bogoliubov (CRHB) calculations with this functional provide a very successful 
description of  different  types of rotational bands  both at low and high spins 
\cite{NL3*,AO.13}. 

  The CRMF equations are solved in the basis of an 
anisotropic three-dimensional harmonic oscillator in Cartesian coordinates characterized 
by the deformation parameters $\beta_0$ and $\gamma$ and oscillator frequency 
$\hbar \omega_0 = 41 A^{-1/3}$ MeV (see Refs.\ \cite{KR.89,AKR.96}).  The truncation 
of the basis is performed in such a way that all states belonging  to the major shells up to 
$N_F=16$ fermionic shells for the Dirac spinors and up to $N_B=20$ bosonic shells for 
the meson fields are taken into account.  This truncation scheme provides sufficient 
numerical accuracy for the physical  observables of interest \cite{AIR.19}. To verify 
that the numerical test of the convergence of the total binding energies as  a  function of 
the number of fermionic shells $N_F$ has been carried out in a similar fashion to that
done in Ref.\ \cite{AIR.19} (see Fig. 4 in this paper and its discussion).   This test 
covered a number of the configurations of interest in the nuclei ranging from $^{14}$Ne 
up to $^{67}$Kr.  The results obtained  for total binding  energies $E(I)$ at a given spin $I$ 
are very similar to those shown in Fig. 4 (a) of  Ref.\ \cite{AIR.19}: the values of $E(I)$ 
calculated with $N_F=16$ and $N_F=20$ differ by no more than several hundred keV.
The impact of the increase of $N_F$ is even smaller on the relative energies of different
configurations (such as those shown in Fig.\ \ref{Ne14-ERLD} below) since the increase 
of $N_F$ leads to additional binding in all configurations. The density distributions 
(similar to those shown in Fig.\ \ref{density} below) obtained with $N_F=16$ and $N_F=20$ 
are also very similar to each other.  The evolution of the single-particle energies as a function 
of $N_F$ for negative energies is very similar to that shown in Fig.\ 4(b) of Ref.\ \cite{AIR.19}:
they stabilize at $N_F\approx 16$. At positive energies, due to the presence of the Coulomb 
barrier the energies of the single-particle states of interest stabilize at lower $N_F$ (typically 
at $N_F\approx 16$) as compared with the situation seen for neutron single-particle states
near neutron drip line which is illustrated in Fig. 4(b) of Ref.\ \cite{AIR.19}. These results
clearly illustrate that the $N_F=16$, $N_B=20$ basis provides a sufficient numerical accuracy 
for systematic studies undertaken in the present paper. 

%%%%%%%%%%%%%%%%%%%%%%%%%%%%%
\subsection{Nucleonic configurations}
\label{conf}
\subsubsection{Basic features.}
\label{conf-basic}
%%%%%%%%%%%%%%%%%%%%%%%%%%%%%
  
     As mentioned in Sec.\ \ref{n-drip-status} the nucleonic configurations 
participating in collective rotation are built either from only bound single-particle 
states or from the combination of bound and quasi-bound single-particle states.    
In the former case one deals with particle-bound rotational bands, while
in the latter one with quasi-bound rotational bands. Note that nucleonic configuration 
ceases to exist when the particle (proton or neutron) is emitted from occupied 
quasi-bound state(s). 

  In the present paper as well in Ref.\ \cite{AIR.19} we consider the transition 
from quasi-bound part of rotational band to particle-bound one: these two parts 
are built on a fixed nucleonic configuration. Note that due to historical reasons 
in all theoretical publications the rotation of the nucleus is considered in terms 
of increasing angular momentum and we follow this tradition in our discussion. 
In contrast, the situation is opposite in experiment since the feeding of rotational 
band takes place at high angular momentum and then the angular momentum 
within the band decreases via the cascade of sequential $\gamma$-ray transitions 
which carry away the angular momentum. Thus, for the type of the bands discussed 
in the present paper and in Ref.\ \cite{AIR.19}  the particle-bound part of rotational 
band will be populated first in experiment and only then the transition to 
quasi-bound part of rotational band will take place.
 
%%%%%%%%%%%%%%%%%%%%%%%%%%%%%%%
\subsubsection{Configuration labelling.}
\label{conf-label}
%%%%%%%%%%%%%%%%%%%%%%%%%%%%%%% 
 
%%%%%%%%%%%%%%%%%%%%%%%%%%%%%%%%%%%%%%%%%%%%%%%%%%%%%%%%
\begin{figure*}[htb]
\centering
\includegraphics*[angle=0,width=5.07cm]{fig-3-a-rev.eps}
\includegraphics*[angle=0,width=5.25cm]{fig-3-b-rev.eps}
\includegraphics*[angle=0,width=5.07cm]{fig-3-c-rev.eps}
\caption{The energies [relative to a smooth liquid drop reference $E_{RLD} = AI(I+1)$]
and neutron and proton quadrupole ($\beta_2$) and triaxial ($\gamma$) deformations  of the  
$[1,1,1,1]\otimes [2,3,3,2]$  occupation block in $^{14}$Ne as a function of angular momentum 
$I$. The deformation parameters 
$\beta_2 = \sqrt{\frac{5}{16\pi}} \frac{4\pi}{3 A R_0^2} \sqrt{ Q_{20}^2 + 2Q_{22}^2}$
and
$\gamma= \arctan{\sqrt{2} \frac{Q_{22}}{Q_{20}}}$
are extracted
from respective quadrupole moments
$Q_{20} = \int d^3r \rho({\vec r})\,(2z^2-x^2-y^2) $ and
$Q_{22} = \int d^3r \rho({\vec r})\,(x^2-y^2)$. Note that
$R_0=1.2 A^{1/3}$.
\label{conf}
}
\end{figure*}
%%%%%%%%%%%%%%%%%%%%%%%%%%%%%%%%%%%%%%%%%%%%%%%%%%%%%%%             

  In the CRMF calculations, the  configurations are specified by the number of particles 
$k_{par}^r$ (where $k=\pi$ or $\nu$ for protons and neutrons, respectively) occupying 
single-particle orbitals of specific parity $par$  and signature $r$ combinations, namely,  
by the occupation block $[\nu_+^+, \nu_+^-, \nu_-^+, \nu_-^-] \otimes [\pi_+^+, \pi_+^-, \pi_-^+, \pi_-^-]$. 
Note that the occupied orbitals are counted from the bottom of nucleonic potential. For 
simplicity we use only the sign for the $r = \pm i$ signature and parity $par=\pm$
in this block. The occupation 
block $[\nu_+^+, \nu_+^-, \nu_-^+, \nu_-^-]  \otimes [\pi_+^+, \pi_+^-, \pi_-^+, \pi_-^-]$ 
defines the lowest in energy  configuration for a given combination of parities and signatures 
of occupied orbitals. This type of configuration definition is common in density functional theoretical
(DFT) calculations without pairing:  it is used not only in the CRMF calculations \cite{KR.89,AKR.96},
but also in the cranked Skyrme DFT calculations \cite{SDFNPXN.12,Yoshida.22}.

  The rotational and deformation properties of extremely deformed bands strongly depend 
 on intruder content of their configurations. Thus, an alternative way of configuration labeling, 
 frequently used in the  literature, is by counting the number of occupied intruder  orbitals 
 \cite{AKR.96,NWJ.89}. This labelling scheme is also used in the present paper since it
 provides important physical insight on the properties of the bands. Thus, the configurations
 are labeled by shorthand labels [$n, p_1p_2$] where $n$ ($p_1p_2$) indicates the number  
 of occupied  intruder neutron (proton) orbitals. However, the meaning of intruder orbitals 
 depends on the mass region. For example,  the configurations $[n,p_1]$ in $^{14}$Ne are labeled 
 by the number  of occupied $N=2$ neutron ($n$) and proton ($p_1$)  intruder orbitals while 
 those in $^{67}$Kr by the  number of the occupied $N=4$ neutron ($n$) intruder orbitals and $N=4$ 
 ($p_1$) and $N=5$ ($p_2$) proton intruder orbitals. The label  $p_2$ is omitted when $p_2=0$.

  Note that such labeling is feasible only in the case when more than 50\% of intruder orbital 
wavefunction belongs to a specific $N$-shell.  However, proton intruder orbitals, which are 
strongly downsloping  as a function of rotational frequency in the routhian diagrams at high 
rotational frequencies (see Fig.\ \ref{routh}), are significantly mixed because of the 
$\Delta N=2$ interaction. For example, at rotational frequency $\Omega_x=3.20$ MeV the 
squared weights $a_N^2$ of the $N=2$, 4, 6, and 8 shells in the structure of the wave 
function of the intruder I2 orbital of the configuration $[0,{\rm M}2]$ are 0.09,  0.12, 0.19 and 
0.21, respectively.  Note that  $\sum_N a_N^2=1.0$ (see Sec. D1 in Ref.\ \cite{AA.08}).  As a 
consequence, there is no single $N$-shell whose contribution to the structure of the wave 
function exceed 50\%.  Such strongly mixed orbitals are called here as M-orbitals 
and the label "M\#" is used in the shorthand $[n,p]$ notation to indicate their presence
and their number \#.  Note that we count total number  \# of negative and positive  parity 
orbitals when we define the number of the M-orbitals.

  The use of such configuration labelling and basic features of calculated configurations 
are illustrated in Fig.\  \ref{conf} on the example $[1,1,1,1] \otimes [2,3,3,2]$ 
occupation block in $^{14}$Ne. Its calculated $E-E_{RLD}$ curve represents the envelope of 
two configurations defined in terms of the occupation of intruder orbitals, namely, the $[0,3]$ 
and $[0,{\rm M2}]$ [see Fig.\ \ref{conf}(a)].  The configuration change from $[0,3]$  to $[0,{\rm M2}]$ 
within this occupation block is due to unpaired band crossing related to the crossing of the 
single-particle orbitals  which takes place at rotational frequency $\Omega \approx 2.6$ MeV 
(see Fig.\ \ref{routh}).  It triggers substantial changes in equilibrium deformations  and their 
evolution with angular  momentum [see Figs. \ref{conf}(b) and (c)].

  The lines of  different styles are used in Fig.\ \ref{conf}  and throughout the paper to discriminate 
between different parts of the $E-E_{RLD}$ curve corresponding to a given occupation block.
In this paper, proton unbound and/or quasi-bound  parts of the rotational bands  
[such as configuration $[0,3]$ in Fig.\ \ref{conf}] are shown by thin solid lines while the proton 
bound ones [such as the 
configuration $[0,{\rm M}2]$ in Fig.\ \ref{conf}] by thick solid  line. Since the calculations are 
performed as a function of rotational frequency,  the solutions  are not available for some 
spin range in the region of band crossing. Thus, we use dashed thin line to connect the 
configurations, defined in terms of intruder orbitals,  within the envelope of the specific 
occupation block (see Fig.\ \ref{conf}).

%%%%%%%%%%%%%%%%%%%%%%%%%%%%%%%%%%%%%%%%%%%%
\section{The birth of proton-bound rotational bands in $^{14}$Ne and $^{67}$Kr nuclei}
\label{Birth-sec}
%%%%%%%%%%%%%%%%%%%%%%%%%%%%%%%%%%%%%%%%%%%%

%%%%%%%%%%%%%%%%%%%%%%%%%%%%%%%%%%%%%%%%
\begin{figure}[ht]
\centering
\includegraphics*[width=7.5cm]{fig-4-a-rev.eps} \\
\includegraphics*[width=7.5cm]{fig-4-b-rev.eps}
\caption{The energies of calculated configurations in $^{14}$Ne and $^{67}$Kr 
relative to a  smooth liquid drop reference $AI(I + 1)$. The numerical values of 
inertia parameter $A $ are indicated on vertical axis. Additional information on
the configurations of these nuclei are displayed in Table \ref{Table-confs}.
\label{Ne14-ERLD}
}
\end{figure}
%%%%%%%%%%%%%%%%%%%%%%%%%%%%%%%%%%%%%%%%

  To illustrate the basic features of rotation in very proton rich nuclei and their dependence 
on  the mass region we consider the examples of the $^{14}$Ne and $^{67}$Kr nuclei. $^{14}$Ne 
is an extremely proton rich nucleus with 10 protons and only 4 neutrons and $^{67}$Kr has 36 
protons and 31 neutrons. In the Ne isotopic chain the last proton bound  even-even nucleus 
is $^{18}$Ne in the RHB calculations at spin $I=0$  which agrees with the  fact that experimentally 
measured $^{16}$Ne is located beyond  the  two-proton drip line  \cite{AARR.14}.   These calculations 
also predict  that the last proton bound Kr  isotope is $^{68}$Kr. This is in agreement with the observation
of two proton radioactivity in $^{67}$Kr \cite{67Kr-2-proton-emitter,WN.18}.

%%%%%%%%%%%%%%%%%%%%%%%%%%%%%%%%%%
\begin{table}[htb] 
\centering
\caption{The configurations calculated in the $^{14}$Ne and $^{67}$Kr nuclei. 
The first column specifies  the occupation blocks of the configurations. 
The second column identifies the configurations in terms of intruder orbitals
corresponding to the respective occupation block: the first/second label 
corresponds to low/high spin configurations.
The total parity $\pi_{tot}$ and 
signature $r_{tot}$ of the occupation block are listed as 
($\pi_{tot}$, $r_{tot}$) in the last column. 
\label{Table-confs}
}
\begin{tabular}{|c|c|c|c|c|c|} 
\hline       
 occupation block  & configurations & ($\pi_{tot}, r_{tot}$)                                 \\  \hline
          1                  &           2           &  3                                            \\  \hline
 \multicolumn{3} {|c|}{$^{14}$Ne  $[N=4, Z=10]$}                              \\ \hline 
%N1 =
 $[1,1,1,1]\otimes [2,2,3,3]$      &  [0,2], [0,M1]             &  $(+,+)$     \\ \hline
%N2 = 
$[1,1,1,1]\otimes [2,3,3,2]$       &  [0,3], [0,M2]             &  $(-,+)$      \\ \hline
% N3 = 
$[1,1,1,1]\otimes [1,3,4,2]$       &  [0,M1], [0,M3]          &  $(+,+)$     \\ \hline
% N4 = 
$[1,1,1,1]\otimes [2,2,4,2]$       &  [0,M1], [0,M2]          &   $(+,-)$     \\ \hline
% N5 = 
$[1,1,1,1]\otimes [2,2,2,4]$       &  [0,M1], [0,M1]          &  $(+,-)$      \\ \hline
% N6 = 
$[1,1,1,1]\otimes [1,2,4,3]$       &  [0,M1], [0,M2]          &  $(-,+)$      \\ \hline
% N7 = 
$[1,1,1,1]\otimes [1,4,3,2]$       &  [0,3], [0,M3]             &  $(-,-)$       \\ \hline
% N8 = 
$[1,1,1,1]\otimes [3,2,3,2]$       &  [0,3], [0,M2]             &  $(-,-)$       \\ \hline
% S1 = 
$[1,1,1,1]\otimes [2,2,3,3]$       &  [0,2], [0,M1]             &  $(+,+)$     \\ \hline
 \multicolumn{3} {|c|}{$^{67}$Kr  $[N=31, Z=36]$}                             \\ \hline 
%N1 = 
$[7,7,9,8]\otimes [8,8,10,10]$   &  [0,2]                        &  $(-,+)$      \\ \hline
% N2 = 
$[8,8,8,7]\otimes [8,9,9,10]$    &  [2,3], [2,31]              &  $(+,-)$      \\ \hline
%N3 = 
$[8,8,8,7]\otimes [9,9,9,9]$      &  [2,4], [2,41]              &  $(-,+)$      \\ \hline
%S1 = 
$[8,9,7,7]\otimes [9,9,9,9]$      &  [3,4], [3,41]              &   $(+,-)$     \\ \hline
% S2 = 
$[8,8,7,8]\otimes [8,9,10,9]$    &  [2,3], [2,31]              &  $(+,-)$      \\ \hline
% S4 = 
$[8,8,8,7]\otimes [10,9,9,8]$    &  [2,5], [21,51]            &  $(+,-)$      \\ \hline
\end{tabular}
\end{table}
%%%%%%%%%%%%%%%%%%%%%%%%%%%%%%%%

   The calculated configurations of the $^{14}$Ne and $^{67}$Kr nuclei are shown in Fig.\ 
\ref{Ne14-ERLD}.  One can see that absolute majority of the configurations in these 
nuclei are proton quasi-bound  at low spin but with increasing spin and 
undergoing band crossing these configurations become proton bound. The microscopic 
origin of this transition is illustrated in Fig.\ \ref{routh} on the example of proton single-particle 
routhian diagram corresponding to the $[1,1,1,1] \otimes [2,3,3,2]$ occupation block which leads 
to the $[0,3]$ and $[0,{\rm M}2]$ configurations (in terms of intruder orbitals) at low and high spins
[see red lines in Fig.\ \ref{Ne14-ERLD}(a)], respectively.  At low and medium rotational  frequencies
$\Omega_x$,  the occupied proton $1/2[220](r=+i)$ and $3/2[211](r=-i)$ orbitals  are located at 
positive energies [see Fig.\  \ref{routh}(a)]. Thus, the $[0,3]$ configuration is proton 
quasi-bound.  However, with increasing  rotational frequency  intruder orbitals I1 and I2 of negative 
and positive  parities become occupied at $\Omega_x \approx 2.9$ MeV leading to the $[0,{\rm M}2]$ 
configuration which becomes proton bound at $\Omega = 3.19$  MeV.  This microscopic mechanism 
of the creation of proton bound  rotational bands from proton quasi-bound ones is equivalent 
to the birth of particle bound rotational bands  in very neutron rich nuclei near neutron drip line 
discussed earlier in Ref.\  \cite{AIR.19}. Note that this mechanism is active in all nuclei studied 
in Sec.\ \ref{Extension-sect} below.

%%%%%%%%%%%%%%%%%%%%%%%%%%%%%%%%%%%%%%%%%%%%%
\begin{figure}[htb]
\centering
\includegraphics*[angle=0,width=4.75cm]{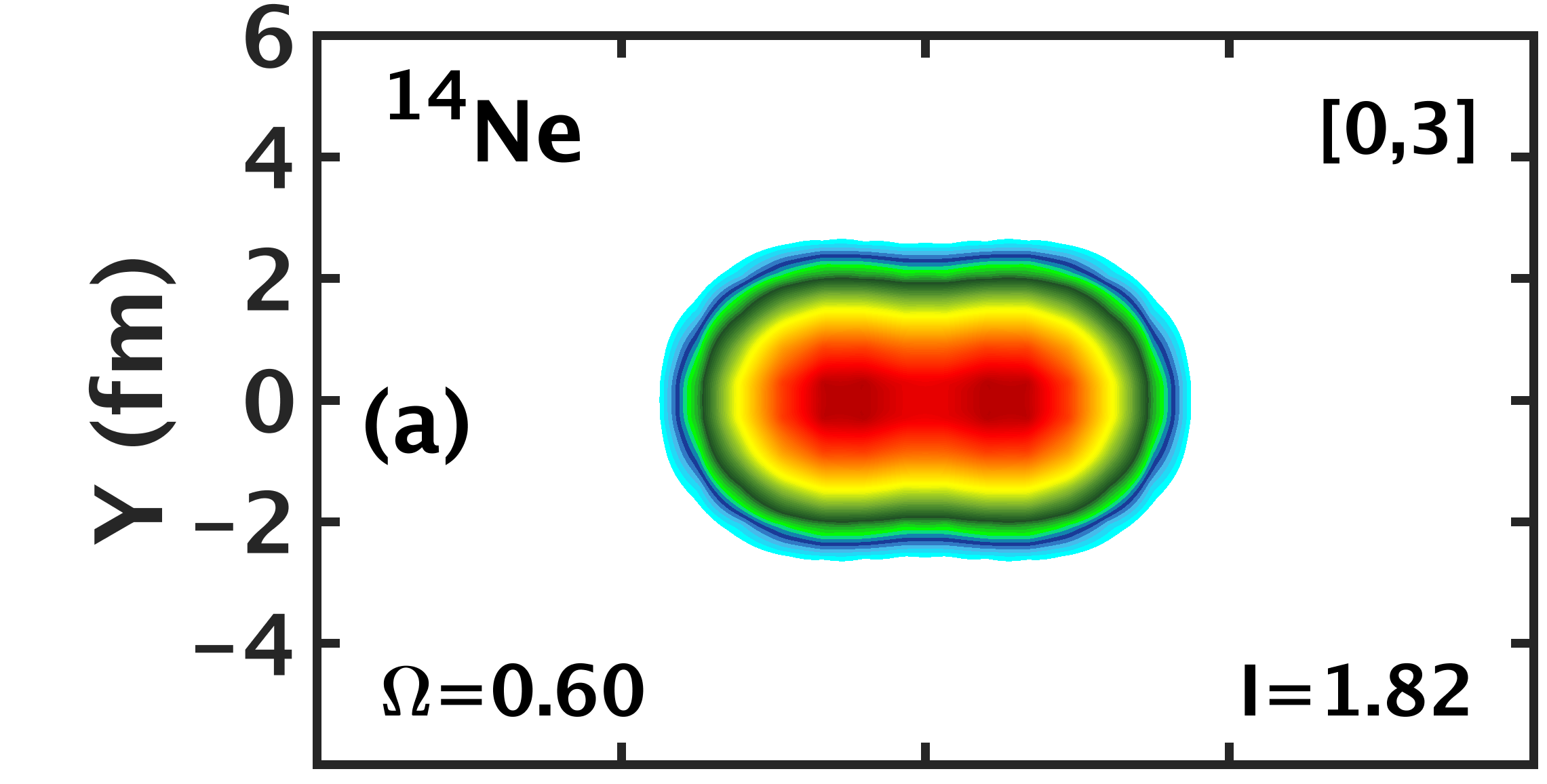}
\includegraphics*[angle=0,width=4.75cm]{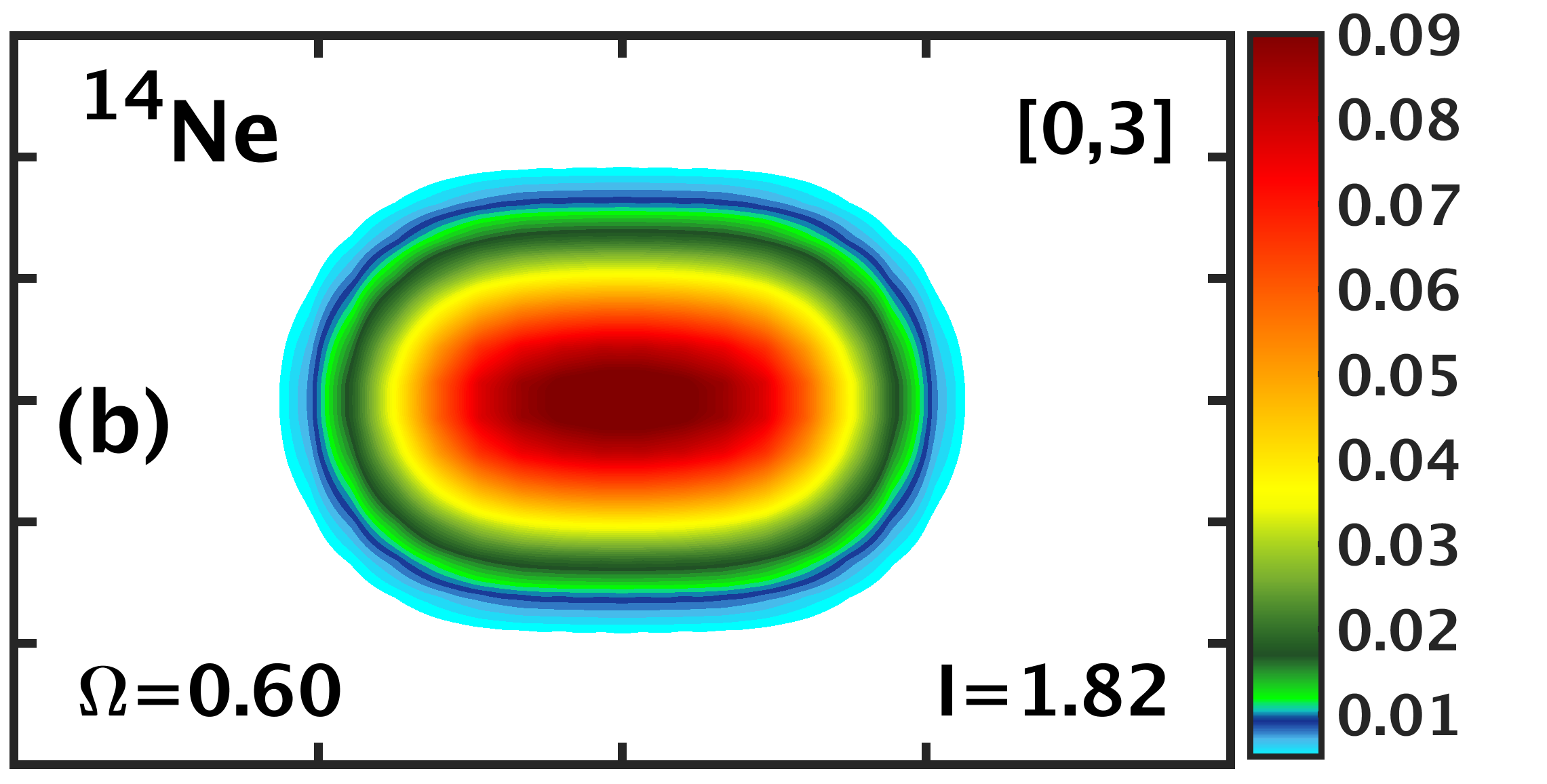} \\
\includegraphics*[angle=0,width=4.75cm]{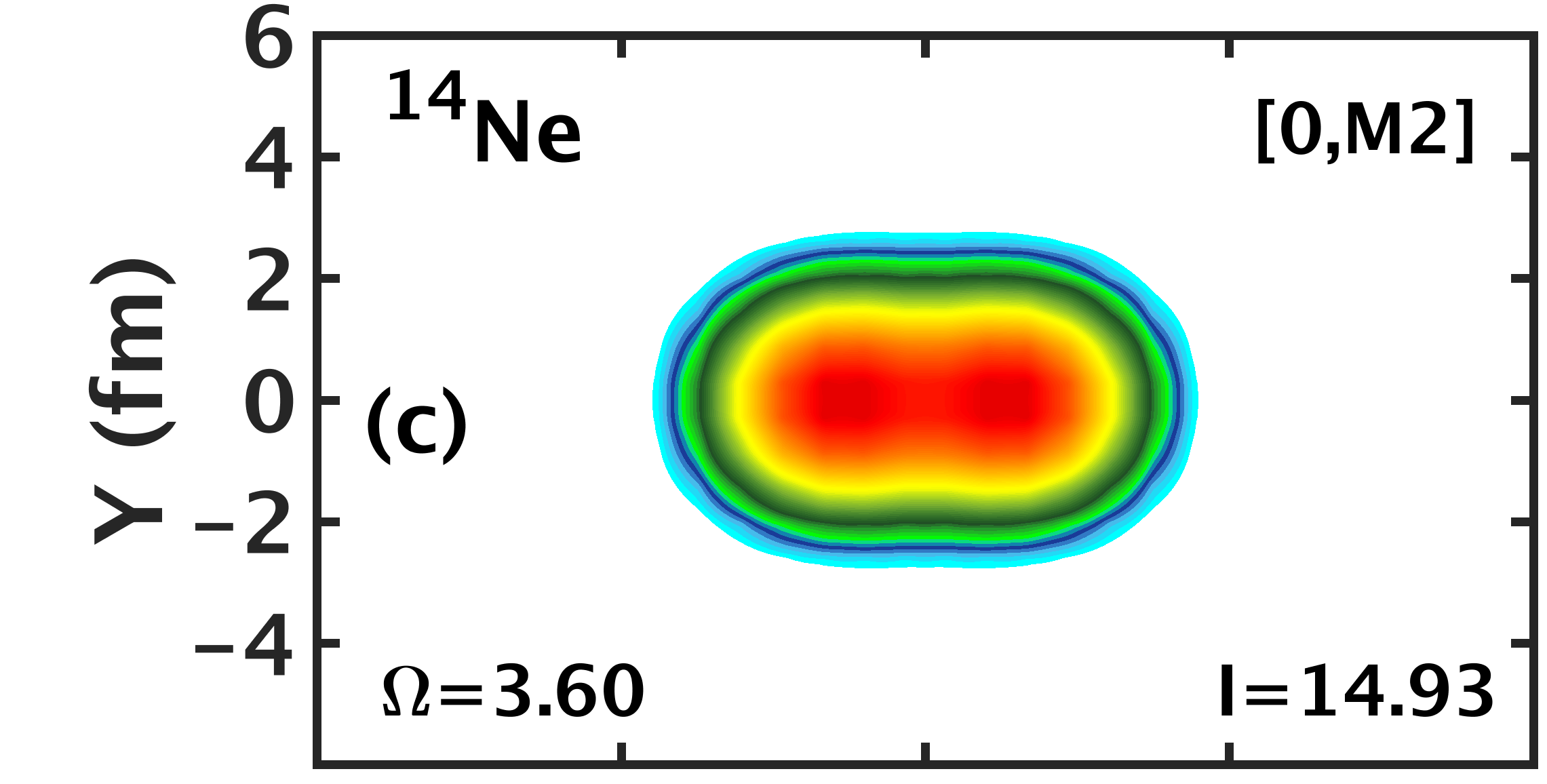}
\includegraphics*[angle=0,width=4.75cm]{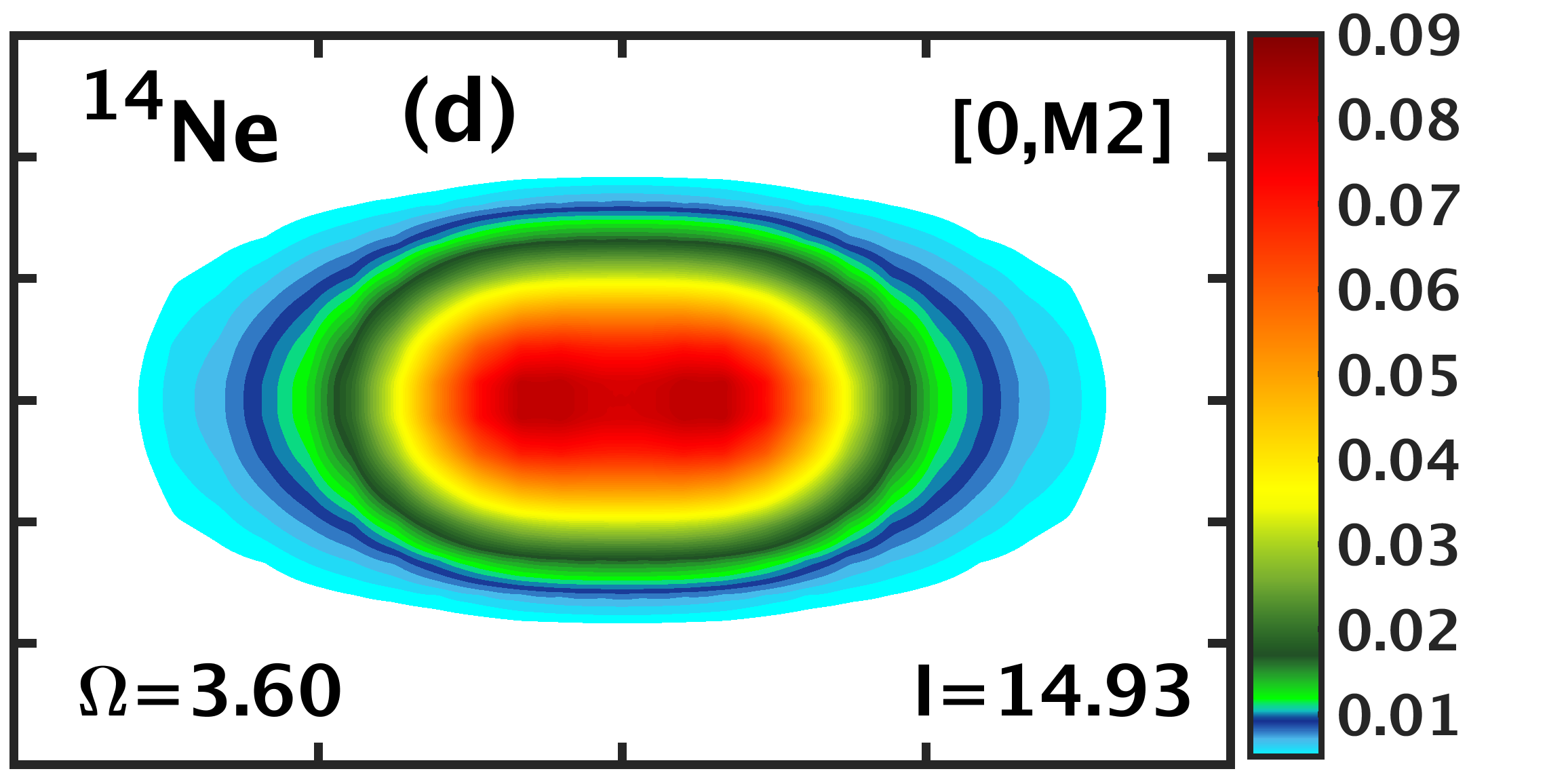} \\
\includegraphics*[angle=0,width=4.75cm]{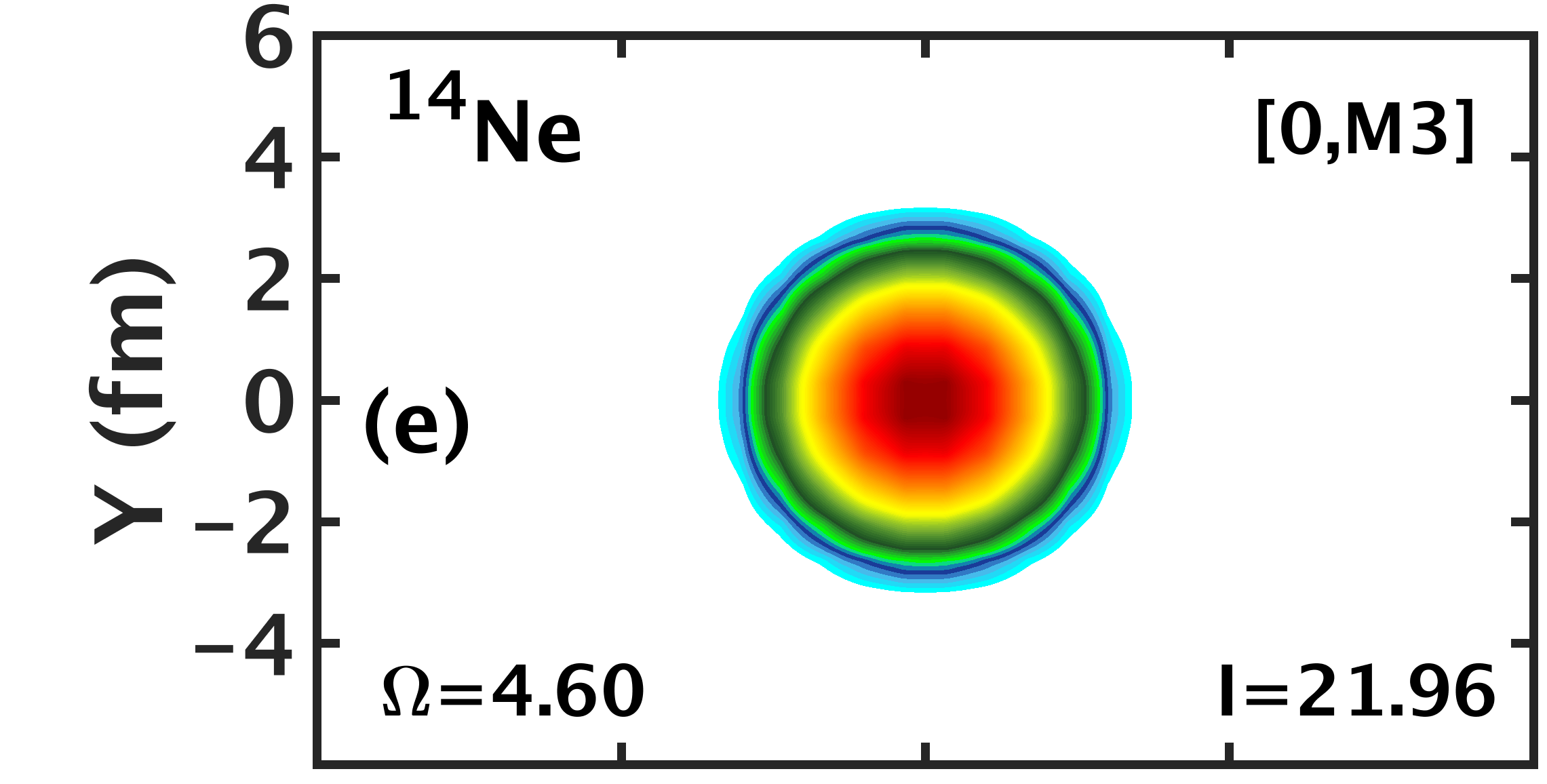}
\includegraphics*[angle=0,width=4.75cm]{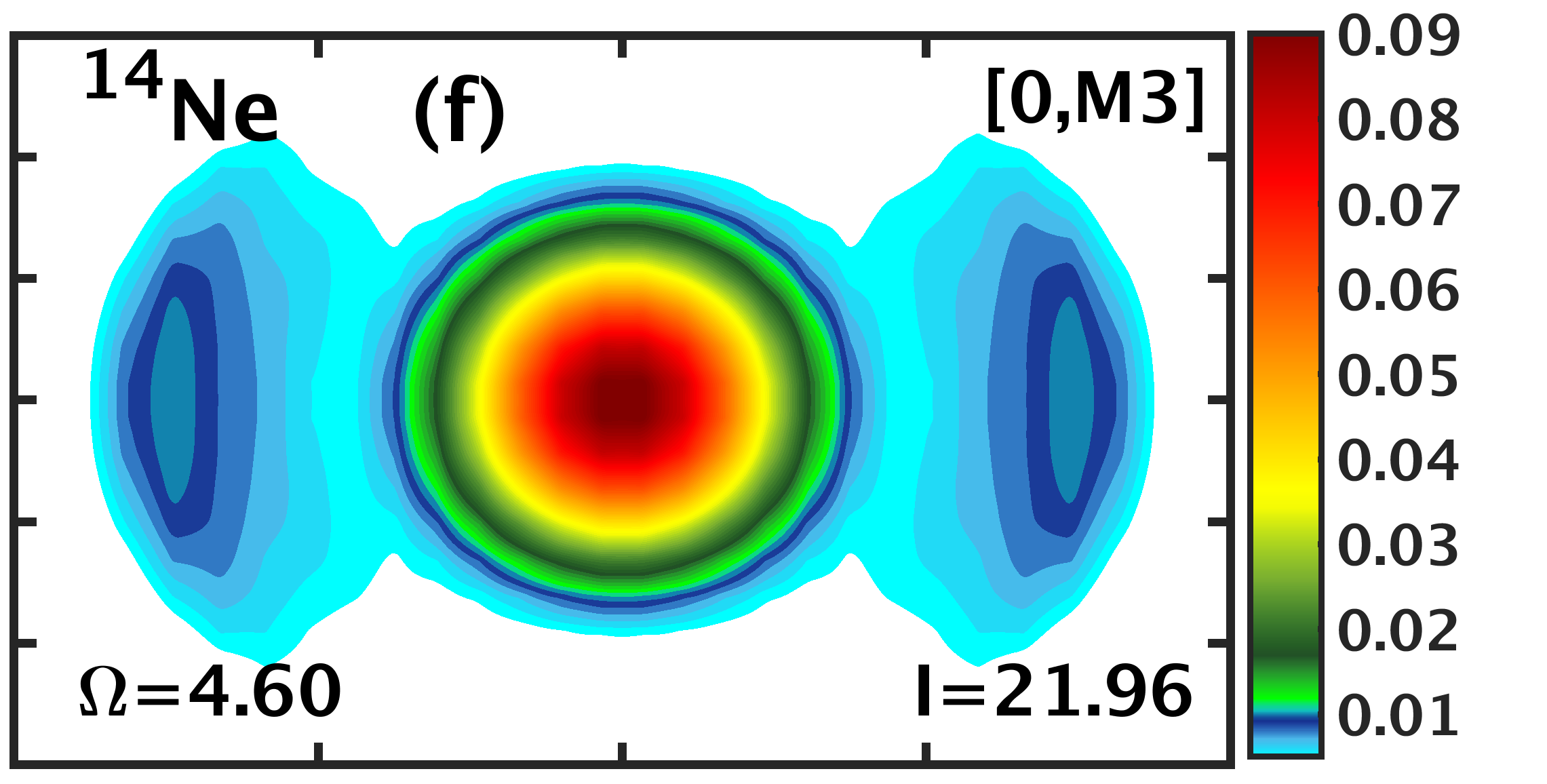} \\
\includegraphics*[angle=0,width=4.75cm]{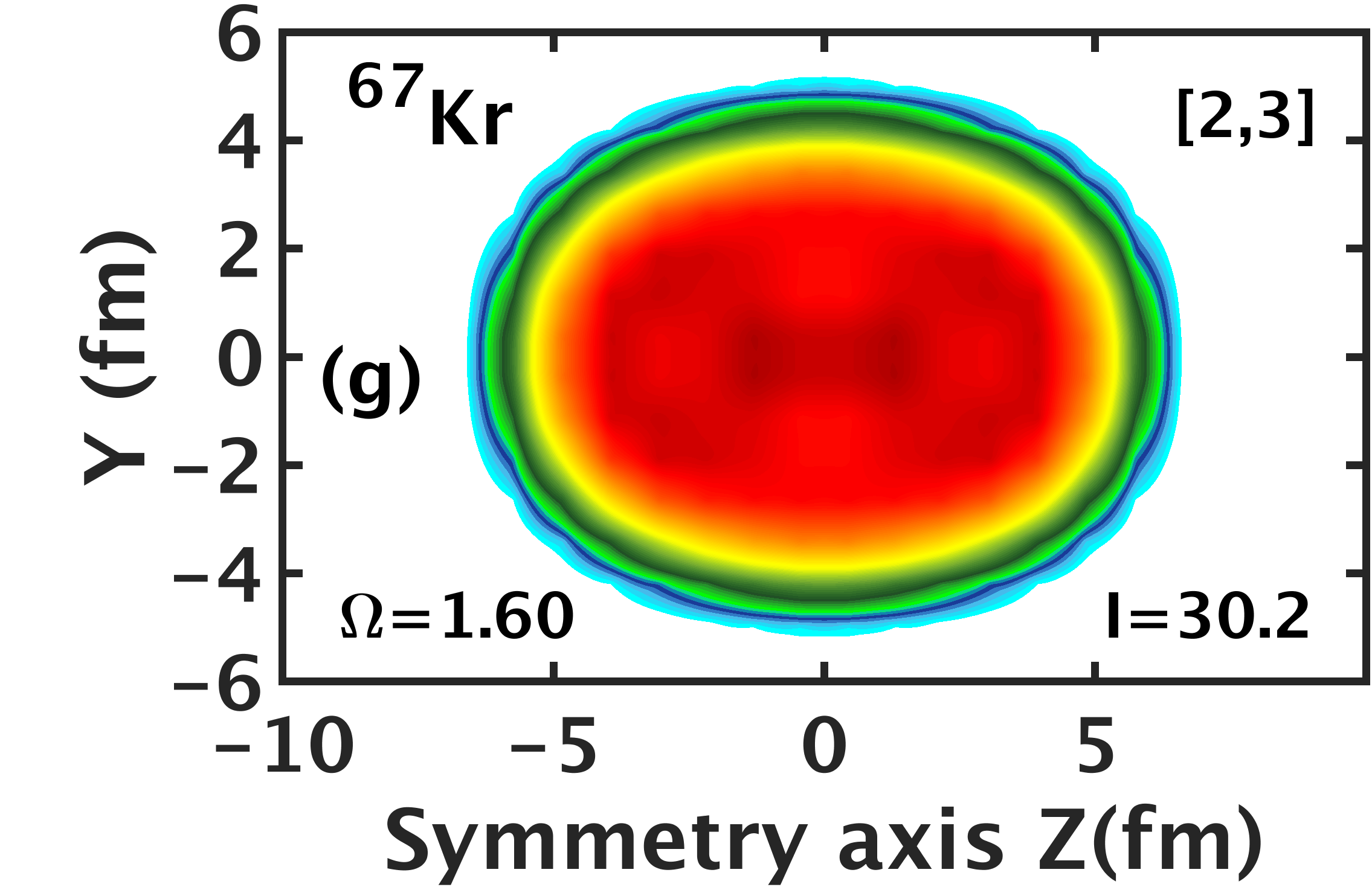}
\includegraphics*[angle=0,width=4.75cm]{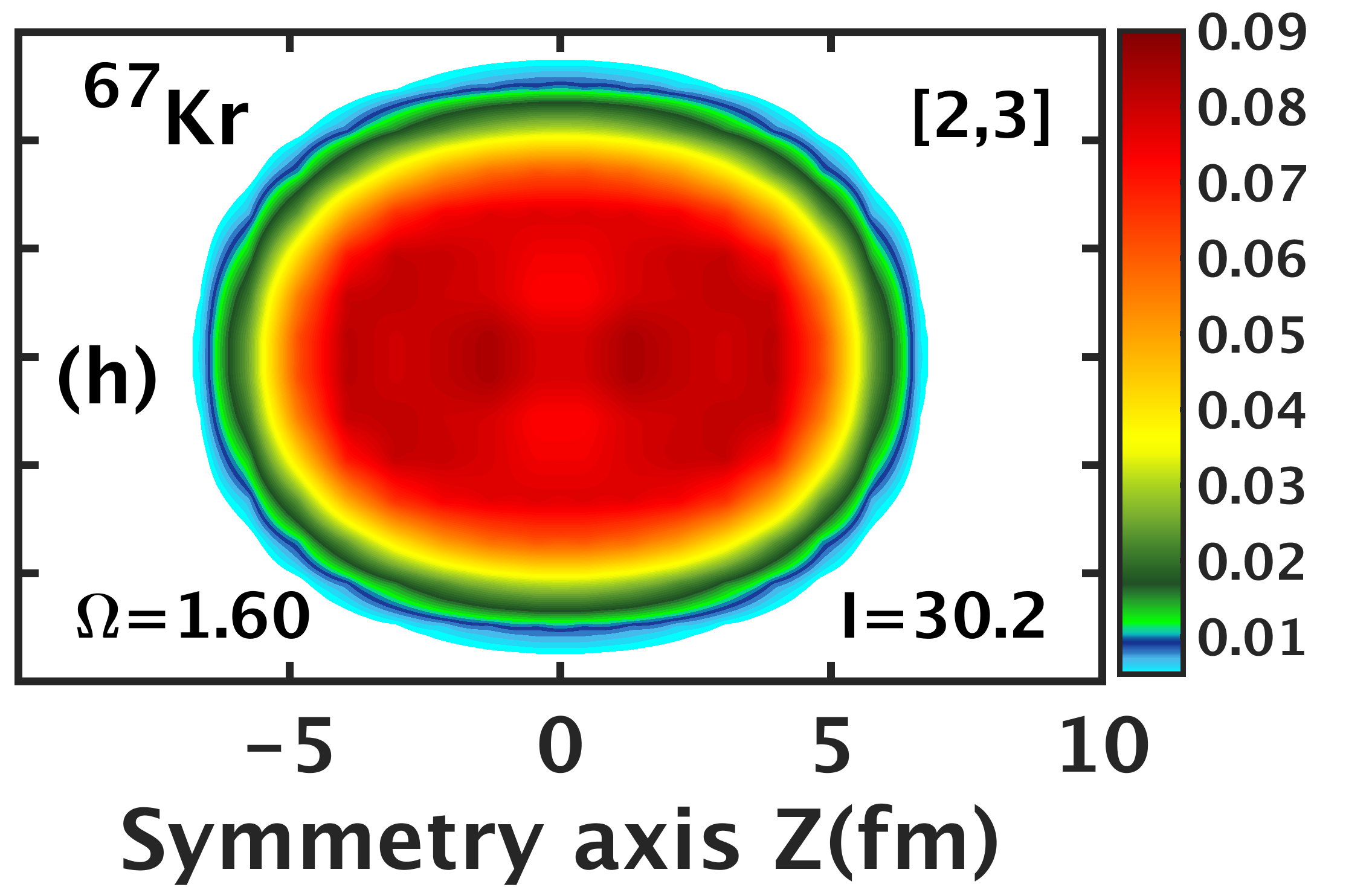}
\caption{
Neutron (left column) and proton (right column) density distributions of 
the configurations in $^{14}$Ne and $^{67}$Kr at indicated  spins and 
frequencies.  The density colormap starts at  $\rho=0.005$ fm$^{-3}$ 
and shows the densities in fm$^{-3}$.
\label{density}
}
\end{figure}
%%%%%%%%%%%%%%%%%%%%%%%%%%%%%%%%%%%%%%%%%%%

   An interesting feature of these rotational structures is the fact that at least 80\% of the  
angular momentum is built in the proton subsystem because of limited angular momentum 
content in the neutron one. Thus, the neutron subsystem behaves as a spectator in building 
collective rotation.

  By performing the calculations for the $^{14}$Ne  nucleus with 
the DD-MEX 
%\cite{TAAR.20} 
and NL1 
\cite{NL1} 
covariant energy density functionals
it was verified  that the general results obtained in the present paper are not affected  by the selection of the  
functional.

%%%%%%%%%%%%%%%%%%%%%%%%%%%%%%%%
\section{Giant proton halo in rotating very proton-rich nuclei}
\label{Halo-sect}
%%%%%%%%%%%%%%%%%%%%%%%%%%%%%%%%

  The above discussed features and transition to exceedingly proton rich nuclei lead to significant
modifications in density distributions which are especially pronounced in light nuclei. For example,
the imbalance between protons and neutrons leads to a substantial proton skin and/or halo already 
at low spin  in all calculated configurations of $^{14}$Ne.  This is illustrated by the comparison of 
neutron and proton densities of the $[0,3]$ configuration in Figs.\ \ref{density}(a) and (b). Additional 
proton particle-hole excitations  leading to the $[0,{\rm M}2]$ configuration generate low density giant 
proton halo which is especially extended in the equatorial region [see Fig.\ \ref{density}(d)]. 
Note that this particle-hole excitation has only small impact on the neutron densities (compare Figs.\ 
\ref{density}(a) and (c))  indicating that there is a substantial decoupling of proton and neutron 
degrees of freedom due to underlying shell structure.

  These two subsystems of the $[0,{\rm M}2]$ 
configuration have drastically different quadrupole ($\beta_2$) and triaxial ($\gamma$) deformations which
also behave differently as a function of angular momentum [see Fig.\ \ref{conf}(b) and (c)]. The 
$\beta_2$ and $\gamma$ deformations of the neutron subsystem almost do not change with 
increasing angular momentum.  In contrast, the quadrupole deformation $\beta_2$ of proton subsystem
drastically increases with increasing angular momentum due to the centrifugal stretching of the 
nucleus [see Fig.\ \ref{conf}(b)]. Similar (but significantly less pronounced) processes have been seen 
before in the calculations of many hyper- and  megadeformed rotational bands in the $A\approx 40$ 
$N\approx Z$ (see Ref.\ \cite{RA.16})  and  $Z=40-58$ (see Ref.\ \cite{AA.08}) regions of nuclear 
chart. The $\gamma$-deformation of proton subsystem decreases fast with increasing angular
momentum. 

   Additional particle-hole excitations lead to more exotic nuclear shapes. For example, the
excitation indicated by blue arrow in Fig.\ \ref{routh} leads to the occupation block  
$[1,1,1,1] \otimes [1,3,4,2]$  which corresponds 
to the envelope of the [0,3] and $[0,{\rm M}3]$ configurations shown in cyan in Fig.\ 
\ref{Ne14-ERLD}. The densities of low spin [0,3]  configuration are similar to those shown in Fig.\ 
\ref{density}(a) and (b). However, the ones for the high spin $[0,{\rm M}3]$ configuration are built 
of near spherical  shape for neutron densities [Fig.\ \ref{density}(e)] and cluster-type giant proton 
halo  in proton subsystem [Fig.\ \ref{density}(f)]. The latter consist of high density 
prolate central  cluster and low (with $\rho \approx 0.01$ fm$^{3}$) 
density clusters centered at around $z\approx \pm 7 $ fm. 

  The kinematic moments  of inertia $J^{(1)}$  and transition quadrupole moments $Q_t$ of the 
bands with giant proton halos are substantially larger than those of the ground state 
rotational band. For example, in the $^{14}$Ne nucleus the kinematic moment of inertia of the 
ground state band is $J^{(1)} \approx 2.2$ MeV$^{-1}$ at spin $I=2\hbar$, while 
that for the [0,M3] configurations shown in Fig. \ref{Ne14-ERLD}(a) is $J^{(1)} \approx 5.3$ 
MeV$^{-1}$ at $I\approx 20\hbar$. The $Q_t$ values for the ground state and
giant proton halo rotational bands are $Q_t \approx 0.55$ $e$b and $Q_t \approx 1.4$ 
$e$b, respectively. These large differences in the physical observables could 
be used  for an experimental identification of potential candidates for giant proton 
halo structures.

  Note that the processes of the creation of proton halos are suppressed in higher $Z$
nuclei due to smaller imbalances between protons and neutrons and reduced relative 
impact of a given single-particle orbital into the buildup of the total densities.  This is 
illustrated by the comparison of the proton and neutron density distributions  in the $[2,3]$ 
configuration of  $^{67}$Kr [Figs.\ \ref{density}(g) and (h)].

  The physical mechanism of the creation of giant  proton halos in rotating nuclei is 
completely different as compared with that predicted for non-rotating proton rich nuclei.  
In the latter ones, it is  based on the occupation of loosely bound $s$ and $p$ orbitals  
\cite{BH.96,23Al-proton-halo.02,JRFG.04}.  In contrast, it is due to  either a significant 
excess of protons over neutrons [compare Figs.\ \ref{density}(a) and (b)] 
or  the occupation of intruder strongly-mixed M-orbitals [Figs. \ref{density}(d) and (f)]. 
The wave functions of the later ones contain significant admixtures from the high $N$ shells
(see example in Sec.\ \ref{conf-label}). Note also that in many cases 
these orbitals are deeply bound: the Fermi levels,  corresponding  to the energy of  the last 
occupied orbital, of the configurations of interest is located in the range between $-4$ to $-2.5$ 
MeV at the highest calculated rotational frequencies (see, for example, Fig.\ \ref{routh}). The 
nodal structure of these intruder orbitals (see discussion in Ref.\ \cite{AA.18}) plays an 
important role in the creation of the halo and cluster type structures of rotating nuclei under 
study.

%%%%%%%%%%%%%%%%%%%%%%%%%%%%%%%%%%%%%%%%%%%%%%%%%%
\begin{figure}[htb]
\centering
\includegraphics*[angle=0,width=9.2cm]{fig-6-a-rev.eps} \\
\includegraphics*[angle=0,width=9.2cm]{fig-6-b-rev.eps}
\caption{Rotation induced extension of the nuclear landscape. 
Open squares show  the nuclei which are proton bound at no rotation 
in the RHB calculations of Ref.\ \cite{AARR.14}: these predictions are 
compared with experimental data  in Sec.VII of Ref.\  \cite{AARR.14}.  
Colored squares, diamonds and circles  show the lowest spin range (in 
step of $5\hbar$) at which proton bound configurations appear in the 
calculations. In panel (a) they are only shown above the lowest spins 
$I_{unpair}$ above which the disappearance of static pairing correlations 
along the yrast line is expected while in panel 
(b) they are displayed at spin values at which the configurations become proton 
bound in the CRMF calculations  (see text for details). Dashed  filling pattern 
is used to indicate the nuclei in which at least one nucleonic configuration with 
rotation induced giant proton halo of the types shown in Figs.\ \ref{density}(d) 
and (e) appear in the calculations.
\label{landscape}
}
\end{figure}
%%%%%%%%%%%%%%%%%%%%%%%%%%%%%%%%%%%%%%%%%%%%%%%%%%

%%%%%%%%%%%%%%%%%%%%%%%%%%%%%%%%%
\section{Rotation induced extension of the nuclear landscape}
\label{Extension-sect}
%%%%%%%%%%%%%%%%%%%%%%%%%%%%%%%%%

  An impact of rotation on the properties  of very proton rich nuclei similar to 
that discussed in Sec.\ \ref{Birth-sec} has been found for even $Z$ isotopic 
chains with  $4\leq Z \leq 36$. 
Here, we focus on rotation induced extension of the nuclear 
landscape presented in Fig.\ \ref{landscape}. Colored squares or circles in this
figure are used for the nuclei which are proton unbound or quasi-bound at no 
rotation, but some configurations  of  which become proton bound when the nucleus rotates.

  The lowest spin ranges (in step of $5\hbar$) at which proton bound configurations 
appear in the calculations are defined in the following way. In Fig.\ \ref{landscape}(a), they are shown
above the lowest spins $I_{unpair}$ for which the disappearance of static pairing correlations is 
expected. The results of the CRHB calculations and the comparison with the ones of 
the CRMF calculations indicates that static pairing disappears at $I_{unpair} \approx 5 \hbar$, 
$I_{unpair} \approx 10\hbar$ and $I_{unpair}\approx 15\hbar$ in the Ne, Ca and Kr nuclei, 
respectively (see Refs.\ \cite{RA.16,AF.05}).  Thus, $I_{unpair}$ is set at 5$\hbar$ for the 
nuclei with $Z\leq 10$ and then gradually increases from $5\hbar$ up to 10$\hbar$  for the nuclei 
with $12\ \leq Z \leq 20$ and from $10\hbar$ up to 15$\hbar$  for the nuclei with $22\leq Z \leq 36$. 
Note that above these spins we consider not only yrast but also excited configurations which are 
located not significantly higher than 2 MeV above the yrast line.  The configurations located at 
higher excitation energies are excluded from the analysis since rotational damping may lead to a 
disappearance of discrete rotational bands \cite{DHLBBMV.96}.

   The existing experience with CRHB and CRMF calculations shows that static pairing correlations 
could disappear due to the combination of blocking and Coriolis anti-pairing effects at spins below 
$I_{unpair}$ when the configurations are based on particle-hole excitations.  Thus, in  Fig.\ \ref{landscape}(b) 
we also show the lowest spins $I_{lowest}$ ($I_{lowest} < I_{unpair}$) at  which such configurations become proton bound 
in the calculations without pairing.  This leads to lowering of the spins as compared with those
given in Fig.\ \ref{landscape}(a) in some nuclei (such as $^{16,17}$Ne) located close to the proton
drip line. The CRHB calculations with pairing are needed to either confirm or reject this lowering 
of the spins but they go beyond the scope of the present paper. Note that for $I<I_{unpair}$ we consider 
only the yrast configurations.

   Fig.\ \ref{landscape} shows that in many nuclei located beyond $I=0$ proton drip line
the transition to proton bound configurations takes place at relatively modest spin values.  However,  
in a given isotopic chain  such transition occurs at higher spin values with increasing  proton excess. 
In addition, the calculations suggest that the formation of rotation induced giant proton halos 
is feasible only in the $Z\leq 20$ nuclei.

   It is necessary to recognize that the rotational excitations in such proton-rich nuclei (with 
considerable excess of the protons over neutrons in light nuclei) have not been experimentally 
studied so far. However, recent experimental studies of the $^{9}$N nucleus indicate the existence
of the quasi-bound (resonance) states (see Refs.\ \cite{CS.23,9-N.PRL.23}) on base of which 
the rotational excitations can be built.  Note that this is extremely proton-rich nucleus which 
consists of seven protons and only two neutrons: the proton/neutron ratio $Z/N$ for this nucleus 
is 3.5 which is substantially larger than the $Z/N=2.5$ value for the $^{14}$Ne nucleus discussed 
in Sec.\ \ref{Birth-sec}. This suggests  that in principle such experimental studies may be feasible 
in the future.

%%%%%%%%%%%%%%%%%%%%%%%%%%%%%%%%%
\section{Conclusions}
\label{concl}
%%%%%%%%%%%%%%%%%%%%%%%%%%%%%%%%%

   The detailed investigation of  fast rotation in the nuclei located in the
vicinity of the drip lines defined at no rotation and beyond has been 
carried out in the cranked relativistic mean field approach. The primary
attention has been paid to very proton rich rotating nuclei. The main 
results of this study can be summarized as follows.

\begin{itemize}

\item
  The analysis of existing theoretical studies of the rotation in the nuclei near 
and beyond drip lines indicates that collective rotation can lead to an 
increase of the stability of nuclear systems via two mechanisms. In the
first one, the particle-hole excitations from the states with low orbital 
momentum $l$ to the ones with high $l$ values lead to an increase
of centrifugal barrier of quasi-bound single-particle states occupied
in the nucleonic configuration. This leads to a significant reduction of 
the decay width of the rotational states of quasi-bound rotational  
bands.  In the second one, with increasing rotational frequency Coriolis 
interaction acting on occupied high-$j$ (high-$l$) orbitals, which are 
quasi-bound at low frequencies, drives them below zero energy 
threshold  and makes them bound.

\item
  For the first time we show that rotational bands which are proton quasi-bound  at 
zero or low spins can be transformed into proton bound ones by collective rotation 
of nuclear systems.  Strong Coriolis interaction acting on high-$j$ or strongly 
mixed M orbitals provides a microscopic mechanism for this transformation by 
driving the highest in energy occupied single-particle states of nucleonic configurations 
into negative energy domain. 
These rotational bands, which in the language of Ref.\ \cite{AIR.19} undergo the birth of 
particle bound parts at high spin, are transitional in nature between classical particle-bound 
rotational bands which  at present are almost exclusively observed in experiment 
\cite{VDS.83,BHN.95,PhysRep-SBT,DHLBBMV.96}  and very rare examples of 
quasi-bound rotational bands \cite{FNJMP.16,FRMLN.16}.

\item 
  The calculations indicate the presence of rotation induced giant proton 
halo in the nucleonic configurations which are typically proton bound at high spin.
Few  of these configurations reveal underlying clustering structures. Their formation 
is triggered by the occupation of strongly mixed M intruder orbitals. Thus, the physical 
mechanism of the creation of giant proton halos in rotating nuclei is completely different 
as compared with non-rotating nuclei in which the formation of proton halo proceeds  
via the occupation of loosely bound $s$ and $p$ orbitals 
\cite{BH.96,23Al-proton-halo.02,JRFG.04}.

\item
The birth of proton bound rotational bands leads to a substantial extension of nuclear 
landscape towards more proton rich nuclei as compared with the one defined in
non-rotating nuclei. This is similar to the extension of nuclear landscape
towards more proton rich nuclei on transition from ellipsoidal to toroidal shapes in 
hyperheavy nuclei \cite{AA.21}. In both cases, the change of collective coordinates 
(deformation in hyperheavy nuclei and combination of deformation and rotational frequency 
in  rotating nuclei) triggers the modification of underlying single-particle structure in such 
a way that occupied states become proton bound above some values of these coordinates. 
Note that  in a given isotopic chain the transition  due to collective rotation to proton bound 
configurations occurs at  higher spin values with increasing  proton excess. 
\end{itemize}

   This material is based upon work supported by the U.S. Department of Energy,  
Office of Science, Office of Nuclear Physics under Award No. DE-SC0013037.
Useful discussions with Profs. A. Volya and 
D. Vretenar are greatly appreciated.

\end{document}